\documentclass[twocolumn]{aa}
\usepackage{graphicx}
\usepackage{txfonts}
\usepackage{psfig}
\usepackage{natbib}

\newcommand{\be}{\begin{equation}}
\newcommand{\ee}{\end{equation}}
\def\ltsima{$\; \buildrel < \over \sim \;$}
\def\lsim{\lower.5ex\hbox{\ltsima}}
\def\gtsima{$\; \buildrel > \over \sim \;$}\def\gsim{\lower.5ex\hbox{\gtsima}}

\newcommand{\XMM}{XMM-\textit{Newton}}
\newcommand{\Swift}{\textit{Swift}}

\begin{document}

\title{\Swift{} and \XMM{} observations of the dark GRB\,050326}

\author{A. Moretti\inst{1} \and A. De Luca\inst{2} \and D. Malesani\inst{3}
\and S. Campana\inst{1} \and A. Tiengo\inst{2} \and A. Cucchiara\inst{4} \and
J.N. Reeves\inst{5} \and G. Chincarini\inst{1,6} \and C. Pagani\inst{1,4} \and
P. Romano\inst{1} \and G. Tagliaferri\inst{1} \and P. Banat\inst{1} \and M.
Capalbi\inst{7} \and M. Perri\inst{7} \and G. Cusumano\inst{8} \and V.
Mangano\inst{8} \and T. Mineo\inst{8} \and V. La Parola\inst{8} \and A.
Beardmore\inst{9} \and M. Goad\inst{9} \and J.P. Osborne\inst{9} \and J.E.
Hill\inst{5,10} \and L. Angelini\inst{5,11} \and D.N. Burrows\inst{4} \and S.
Kobayashi\inst{4} \and P. M\'esz\'aros\inst{4} \and B. Zhang\inst{12} \and S.D.
Barthelmy\inst{5} \and L. Barbier\inst{5} \and N.E. White\inst{5} \and E.E.
Fenimore\inst{13} \and L.R. Cominsky\inst{14} \and N. Gehrels\inst{5}}

\offprints{A. Moretti; \email{moretti@merate.mi.astro.it}}

\institute{ 
INAF, Osservatorio Astronomico di Brera, via E. Bianchi 46, I-23807 Merate (LC), Italy
\and        
INAF, Istituto di Astrofisica Spaziale e Fisica Cosmica di Milano, via E. Bassini 15, I-20133 Milano Italy
\and        
International School for Advanced Studies (SISSA/ISAS), via Beirut 2-4, I-34014 Trieste, Italy
\and        
Department of Astronomy and Astrophysics, Pennsylvania State University, 525 Davey Lab, University Park, PA 16802, USA
\and        
NASA/Goddard Space Flight Center, Greenbelt Road, Greenbelt, MD20771, USA
\and        
Universit\`a degli studi di Milano-Bicocca, Dipartimento di Fisica, piazza delle Scienze 3, I-20126 Milano, Italy
\and        
ASI Science Data Center, via G. Galilei, I-00044 Frascati (Roma), Italy
\and        
INAF, Istituto di Astrofisica Spaziale e Fisica Cosmica di Palermo, via U. La Malfa 153, I-90146 Palermo, Italy 
\and        
X-Ray Observational Astronomy Group, Department of Physics and Astronomy, University of Leicester, LE1 7RH, UK
\and        
Universities Space Research Association, 10211 Wincopin Circle, Suite 500, Columbia, MD, 21044-3432, USA
\and        
Department of Physics and Astronomy, The Johns Hopkins University, 3400 North Charles Street, Baltimore, MD 21218, USA
\and        
Department Physics, University of Nevada, Las Vegas, Nevada, 89154-4002, USA
\and        
Los Alamos National Laboratory, Los Alamos, New Mexico, 87545, USA
\and        
Department of Physics and Astronomy, Sonoma State University, Rohnert Park, California 94928, USA
}

\date{Received ; accepted }
\titlerunning{\Swift{} and \XMM{} observations of GRB\,050326}
\authorrunning{Moretti et al.}

\abstract{We present \Swift{} and \XMM{} observations of the bright gamma-ray
burst GRB\,050326, detected by the \Swift{} Burst Alert Telescope. The \Swift{}
X-Ray Telescope (XRT) and \XMM{} discovered and the X-ray afterglow
beginning 54~min and 8.5~hr after the burst, respectively. The prompt
GRB\,050326 fluence was $(7.7\pm0.9) \times 10^{-6}$~erg~cm$^{-2}$
(20--150~keV), and its spectrum was hard, with a power law photon index $\Gamma
= 1.25 \pm 0.03$. The X-ray afterglow was quite bright, with a flux of $7
\times 10^{-11}$~erg~cm$^{-2}$~s$^{-1}$ (0.3--8~keV), 1~hr after the burst. Its
light curve did not show any break nor flares between $\sim 1$~hr and $\sim
6$~d after the burst, and decayed with a slope $\alpha = 1.70\pm0.05$. The
afterglow spectrum is well fitted by a power-law model, suffering absorption
both in the Milky Way and in the host galaxy. The rest-frame Hydrogen column
density is significant, $N_{{\rm H},z} \ga 4 \times 10^{21}$~cm$^{-2}$, and the
redshift of the absorber was constrained to be $z > 1.5$. There was good
agreement between the spatial, temporal, and spectral parameters as derived by
\Swift-XRT and \XMM. By comparing the prompt and afterglow fluxes, we found
that an early break probably occurred before the beginning of the XRT
observation, similarly to many other cases observed by \Swift. However, the
properties of the GRB\,050326 afterglow are well described by a spherical
fireball expanding in a uniform external medium, so a further { steepening}
is expected at later times. The lack of such { a} break allowed us to
constrain the jet half-opening angle $\vartheta_{\rm j} \ga 7\degr$. Using the
redshift constraints provided by the X-ray analysis, we also estimated that the
beaming-corrected gamma-ray energy was larger than $3 \times 10^{51}$~erg, at
the high end of GRB energies. Despite the brightness in X rays, only deep
limits could be placed by \Swift-UVOT at optical and ultraviolet wavelengths.
Thus, this GRB was a ``truly dark'' event, with the optical-to-X-ray spectrum
violating the synchrotron limit. The optical and X-ray observations are
therefore consistent either with an absorbed event or with a high-redshift one.
To obey the Ghirlanda relation, a moderate/large redshift $z \ga 4.5$ is
required.
}

\maketitle

\section{Introduction}\label{sec:intro}

The \Swift{} satellite \citep{Gehrels04} is a mission dedicated to the study
of gamma-ray bursts (GRBs) and their afterglows. GRBs are detected and
localized by the Burst Alert Telescope \citep[BAT;][]{Barthelmy05}, and
followed up at X-ray (0.2--10 keV) and optical/ultraviolet (1700--6000~\AA)
wavelengths by the X-Ray Telescope \citep[XRT;][]{Burrows_XRT} and the
Ultraviolet/Optical Telescope \citep[UVOT;][]{Roming05}. During the first year
of operation, \Swift{} has observed some 75 GRB afterglows, already doubling
the pre-\Swift{} sample. This rich dataset has allowed to study in detail the
X-ray light curves, both at early and late times, leading to the discovery of a
complex behaviour \citep[e.g][]{Taglia05,Nousek05,Chinca05,Cusu_050319}.
Coupled with optical data, either from UVOT \citep[e.g.][]{Blustin05} or
ground-based observatories \citep[e.g.][]{Berger05}, this has opened a new era
in the afterglow modeling. \Swift{} also provided the first detection of truly
dark GRBs, that is, events with no optical emission up to very deep limits
\citep{Roming_dark}. The study of high-redshift GRBs has also started, with the
discovery of the first burst at $z > 6$
\citep{Watson05b,Cusu_050904,Haislip05,Price05,Taglia_050904,Kawai05}.
Moreover, it was found that \Swift{} GRBs have a larger average redshift than
those discovered by earlier missions \citep{Jakob05}.

During the performance verification and calibration phase (2004 Nov 20 through
2005 Apr 5), \Swift{} observed sixteen GRB afterglows. Twelve of them were
observed in automatic mode, and, among these, eight could be promptly (within
200~s since the trigger) observed by XRT and UVOT. In the remaining four cases,
the beginning of the observation was delayed by approximately 50~min due to
the  Earth occultation constraints. This is the case for the bright
GRB\,050326, which was discovered by BAT on 2005 Mar 26 at 9:53:55 UT
\citep{Markwardt05}. Its coordinates were $\alpha_{\rm J2000} = 00^{\rm
h}27^{\rm m} 34^{\rm s}$, $\delta_{\rm J2000} = -71\degr22\arcmin34\arcsec$,
with an uncertainty radius of 3\arcmin{} \citep[95\%
containment;][]{Cummings05}. This burst was also detected by the
\textit{Wind}-Konus experiment \citep{Golenetskii05}, leading to the
characterization of its broad-band gamma-ray spectrum.

The \Swift{} narrow field instruments could begin observing only 54~min after
the BAT trigger. A bright, uncatalogued X-ray source was detected by XRT inside
the BAT error circle, and was proposed to be the X-ray afterglow
\citep{Moretti05}. However, no source was detected by UVOT at this location
\citep{Holland05}. XRT collected data up to 6.15~d after the burst.
Subsequently, the decay of the light curve prevented any further detection
of the afterglow. This object was also observed for 45.8~ks by \XMM{}
\citep{EhlePM05,DeLuca_GCN}, starting 8.5~hr after the trigger.

Only limited ground-based follow-up was reported for this burst. This was
likely due to its unfavorable location in the sky (very few telescopes can
point at such low declination), as well as to the brightness of the Moon (which
was 99\% full at the time of the GRB explosion). No counterpart at wavelengths
other than the X rays was reported.

In this work, we present a complete discussion of the \Swift{} and \XMM{} observations of
GRB\,050326. In Sect.~\ref{sec:prompt} we describe the properties of the prompt
emission. In Sect.~\ref{sec:XRT} we describe in detail the XRT observations, the
data reduction procedure, and the temporal and spectral analysis; in
Sect.~\ref{sec:XMM} we do the same for the \XMM{} data. In
Sect.~\ref{sec:XRT_XMM} we compare the results of the two instruments. In
Sect.~\ref{sec:UVOT} we describe the UVOT optical observations. Finally, in
Sect.~\ref{sec:discussion} we present the physical implications of our
observations in the framework of the standard GRB afterglow model. Our
conclusions are summarized in Sect.~\ref{sec:conclusion}.

Throughout this paper, all errors will be quoted at 90\% confidence level for
one parameter of interest, unless otherwise specified. The reduced $\chi^2$
will be denoted as $\chi^2_\nu$, and the number of degrees of freedom with the
abbreviation ``d.o.f.''. We follow the convention $F_\nu(\nu,t) \propto
t^{-\alpha}\nu^{-\beta}$, where $\alpha$ and $\beta$ are the temporal decay
slope and the spectral index, respectively. As time origin, we will adopt the
BAT trigger \citep{Markwardt05}. The photon index is $\Gamma = 1 + \beta$.
Last, we adopt the standard ``concordance'' cosmology parameters, namely
$\Omega_{\rm m} = 0.27$, $\Omega_\Lambda = 0.73$, $h_0 = 0.71$
\citep[e.g.][]{Spergel03}.

\section{Prompt emission}\label{sec:prompt}

\begin{figure}
\includegraphics[width=\columnwidth]{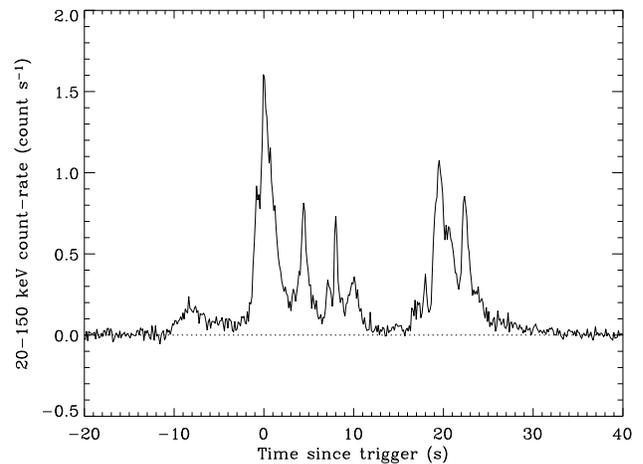}

\caption{The background-subtracted BAT light curve in the 20--150~keV energy
band. The origin of the time axis was set to the instrument trigger, but a weak
peak is apparent $\approx 9$~s before.\label{fg:BAT_lc} }

\end{figure}

We reduced the BAT data using the latest available release of the HEADAS
software (version 1.4). The light curve in the BAT energy band (20--150~keV)
presents an initial, weak peak, 9~s before the trigger, followed by several
bright distinct peaks (Fig.~\ref{fg:BAT_lc}). The $T_{90}$ and $T_{50}$
durations of the burst (that is, the time intervals in which 90\% and 50\% of
the fluence were collected, respectively) were 29.5 and 19.3~s, in the 20--150
keV band, respectively. We first modeled the BAT spectrum as a single power law
with photon index $\Gamma$. This provided a good fit ($\chi^2_\nu = 1.06$ for
53 d.o.f.), yielding $\Gamma = 1.25 \pm 0.03$ in the 20--150~keV energy range.
The fluence in the same band was ($7.7 \pm 0.9) \times 10^{-6}$~erg~cm$^{-2}$.

GRB\,050326 also triggered the \textit{Wind}-Konus detector
\citep{Golenetskii05}: in the \mbox{20 keV--3 MeV} energy range it lasted 38~s,
and had a fluence of $(3.22\pm0.05) \times 10^{-5}$~erg~cm$^{-2}$.
\citet{Golenetskii05} fitted the time-integrated spectrum of the burst as
measured by the \textit{Wind}-Konus detector with a Band model \citep{Band93},
that is a smoothly joined broken power law with low- and high-energy photon
indices $\Gamma_1$ and $\Gamma_2$, respectively, and break energy $E_0$. The
best fit provided $\Gamma_1 = 0.74 \pm 0.09$, $\Gamma_2 = 2.49 \pm 0.16$, and
$E_0 = 160 \pm 22$~keV. The corresponding observed peak energy (that is, the
energy at which the maximum of the emission is reached) was $E_{\rm p,obs} =
(2-\Gamma_1)E_0 = 200 \pm 30$~keV. Motivated by their results, we also
performed a fit to the BAT data using the Band function. Since the break energy
$E_0$ lies close to the upper boundary of the BAT energy range (150~keV), we
were forced to freeze $E_0$ and $\Gamma_2$ to the values determined by
\textit{Wind}-Konus. The fit was again good ($\chi^2_\nu = 1.14$ for 53
d.o.f.), and provided $\Gamma_1 = 0.87 \pm 0.03$, in good agreement with the
value found by \citet{Golenetskii05}. It is not surprising that both functional
forms provide a good fit to the data, since they do not differ significantly
inside the BAT energy range. Nevertheless, thanks to the very broad band
covered by the \textit{Wind}-Konus instrument, for this burst the break energy
could be clearly constrained. In the following, we will consider the Band model
as the best description of the GRB\,050326 spectrum. With this fit, the fluence
in the 20--150 keV band was $(7.6 \pm 0.8) \times 10^{-6}$~erg~cm$^{-2}$.
Integrating the burst spectrum from 1 to 10\,000~keV, we could evaluate the
bolometric fluence $\mathcal{F}$ of the burst, finding $\mathcal{F} = 2.4
\times 10^{-5}$~erg~cm$^{-2}$.

No spectral evolution could be detected in the BAT data. We splitted the
observation in three time intervals, covering the ranges $[-9,-1]$, $[-1,13]$,
and $[13,29]$~s (relative to the BAT trigger). By fitting the data with either
a simple power law or with the Band model, the resulting parameters were always
consistent with those derived by fitting the whole spectrum.

\section{XRT data analysis and results}\label{sec:XRT}

\begin{figure}
\includegraphics[width=\columnwidth]{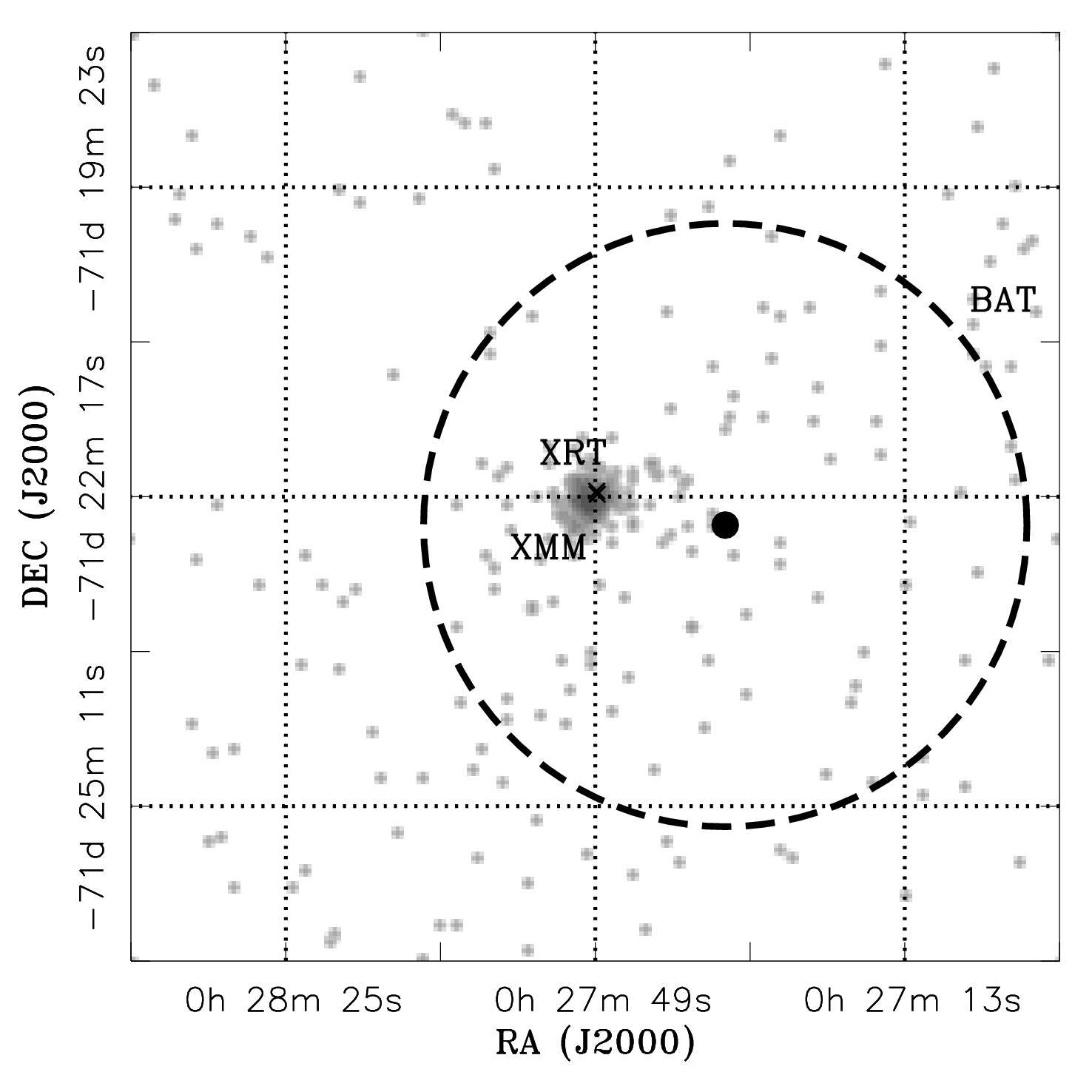}

\caption{XRT image of the field of GRB\,050326, smoothed with a Gaussian
kernel (4\farcs7 full width half maximum). Events were accumulated from the
first segment of the observation (589~s exposure time). The BAT refined
position (black dot) is also shown together with its 95\%
containment error circle (3\arcmin{} radius, dashed line). The XRT and \XMM{}
positions (1\farcs7 apart) are also plotted (crosses), and are almost
indistinguishable. The X-ray error circles (3\farcs6 and 1\farcs5 radius,
respectively) are too small to be { seen} with this scale.\label{imaps}}

\end{figure}

\subsection{Data reduction}\label{sec:XRT_reduction}

For a technical description of XRT and its operations, we refer, e.g., to
\citet{Burrows_XRT} and \citet{Hill04}. XRT started observing the field of
GRB\,050326 on 2005 Mar 26 at 10:48:27 UT, that is, 3307~s after the BAT
trigger. The last observation ended on 2005 Apr 1 at 13:30:53, i.e. 6.15~d
after the burst. Occasionally, some reflected light from the Earth limb made
the very low energy ($< 0.2$~keV) background increase abnormously, so that XRT
incorrectly switched from the photon counting (PC) to the windowed timing (WT)
mode, even if the target count rate was well below 1~count~s$^{-1}$. The
effective exposure time was 59.6~ks in PC mode and 20.7~ ks in WT mode,
leading to the collection of 614 and 580 photons, respectively (0.3-10 keV
energy band). As the satellite settled on the target, XRT recorded a source
count rate of 1.3~count~s$^{-1}$,
which dropped to $3 \times 10^{-4}$~count~s$^{-1}$ at the end of the observing
campaign (2005 Apr 1).
From the third orbit after the start of the observation onwards, the source
count rate was $< 0.1$~count~s$^{-1}$, while the background level was typically
$> 3$~count~s$^{-1}$ over the whole field of view. Since WT data have only one
dimensional spatial information, their S/N ratio was  much lower than that of
PC mode data. We therefore decided to consider WT data  only for the first two
orbits, when the source S/N was higher. Data were reduced using the
\texttt{xrtpipeline} task of the latest available release of the HEADAS
software (version 1.4). Accumulating the PC data from all observations, we
found that the centroid position of the afterglow had coordinates $\alpha_{\rm
J2000} = 00^{\rm h}27^{\rm m}49\farcs16$, $\delta_{\rm J2000} =
-71\degr22\arcmin14\farcs6$, with a 3\farcs6 uncertainty radius (95\%
containment, Fig.~\ref{imaps}).
This position takes into account the correction for the misalignment between
the telescope and the satellite optical axis \citep{Moretti_positions}. This
position is 1\farcm3 away from the refined BAT position \citep{Cummings05}, and
3\farcs4 away from the preliminary XRT position \citep{Moretti05}, calculated
using only the data from the first orbit and without the misalignment
correction.

\subsection{Temporal analysis}\label{sec:XRT_temporal}

\begin{figure*} 
\centering\includegraphics[width=0.8\textwidth]{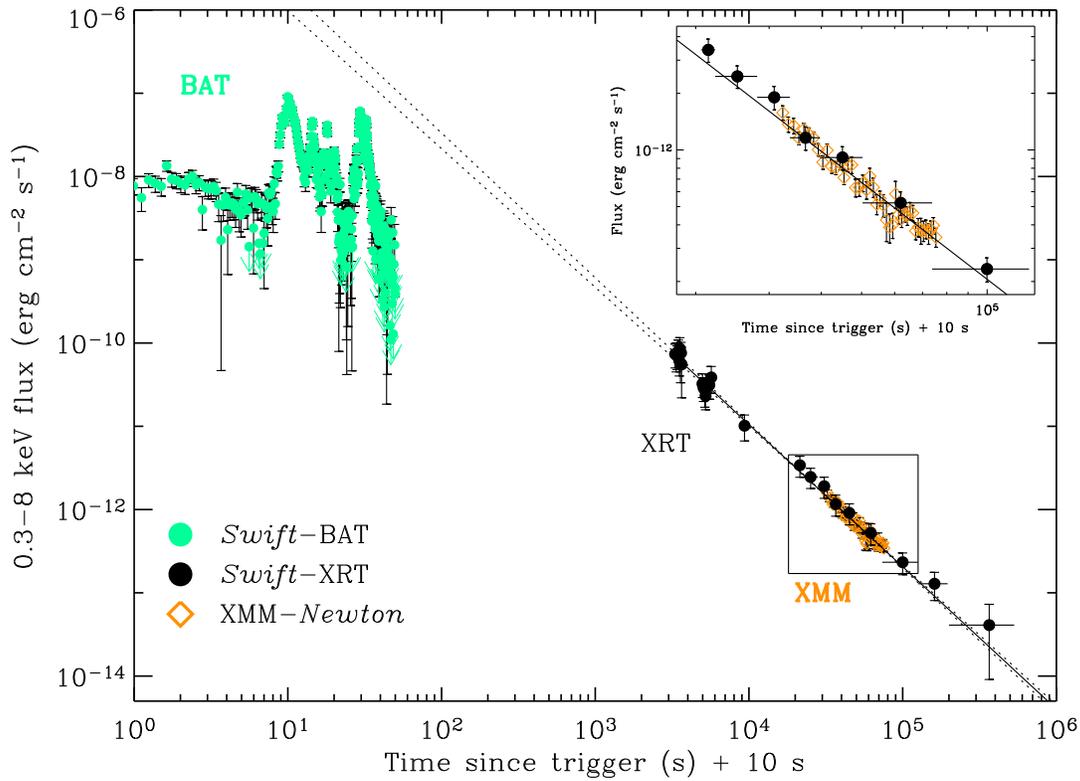} 

\caption{The light curve of GRB\,050326 and of its afterglow in the 0.3--8~keV
energy band (see text for the computation of the flux conversion factors). XRT
(black circles) and \XMM{} data (empty diamonds) show a very good agreement
(see also the inset). The solid line shows the fit to the combined XRT/XMM
afterglow light curve. The dotted lines indicate the 90\% errors of the
extrapolated X-ray light curve.
Light filled circles indicate the extrapolation of the BAT data to the
0.3--8~keV energy range, assuming the Band model as the best-fit spectrum. In
this figure, the time origin was set 10~s before the nominal trigger time, to
show the weak, untriggered precursor. This has no effect on the determination
of the afterglow decay slope, due to the late beginning of the XRT
observation.\label{fig:compa}}

\end{figure*}

In order to extract the light curve, we considered all PC data, but discarded
the WT data taken after the second orbit of the XRT observation ($t > 20$~ks).
PC events were selected having grades 0--12 from a circle with 20 pixel radius
(47\arcsec), corresponding to 92\% of the encircled energy fraction (EEF) at
1.5~keV \citep{Moretti_SPIE}. Only the data in the 0.3--10 keV band energy
range were considered (even if there are no events above 7~keV). To take into
account the pile-up effect, during the initial part of the first orbit ($t \la
4\,000$~s) an annular extraction region with inner radius of 3~pixels
(7\arcsec) was adopted for PC data. This area includes 40\% of the EEF, and the
deriving PSF losses were consequently taken into account.
The accuracy of the PSF model in its central part is $\sim 5$\%
\citep{Moretti_SPIE}. This error was properly propagated when evaluating the
final uncertainty of the PSF-corrected points in the light curve. The
background in PC mode was evaluated by integrating the signal from an annulus
with inner and outer radii of 50 and 90 pixels, respectively, centered at the
afterglow position. Inside this region, the contamination from the afterglow is
expected to be negligible. In WT mode, events were selected having grades 0--2
from a 20 pixel (47\arcsec) wide rectangular region, centered on the detector
X coordinate of the { afterglow}. To estimate the background in WT mode,
we considered a region of the same size centered 40 pixels (94\arcsec) away
from the center of the { afterglow}.

The XRT observation was split in different time segments because of the Earth
occultation constraints. Each satellite orbit lasts $\approx 5\,800$~s, while
the target could be typically observed for approximately 1\,000~s per orbit. To
extract the light curve, the source events were binned in 10~s intervals, and
these bins were further grouped to ensure a minimum of 50 counts per bin. When
the counts in the last bin of each orbit were less than half of the required
minimum (25 counts), the bin was merged with the previous one. From the fifth
orbit onwards, XRT did not collect enough photons within a single orbit, so
data from different orbits were merged. We eventually obtained a
background-subtracted light curve composed by 25 points, with a minimum of 10
and a maximum of 57 counts per bin.

The resulting light curve is shown in Fig.~\ref{fig:compa}, and displays a
uniform decay rate, with no indications of breaks or flares. A single power law
fit provides a good description to the data, yielding a decay slope $\alpha =
1.64 \pm 0.07$. In order to look for spectral variations across the
observation, we computed the afterglow hardness ratio as a function of time. To
this { end}, we selected the events with energy below and above several
pivotal energies, and computed their number ratio. No significant variation was
found during the whole observation, after setting the pivotal energy to 1, 1.5
and 2~keV.

\subsection{Spectral analysis}\label{sec:XRT_spectral}

\begin{figure*}
\centering\includegraphics[angle=270,width=0.8\textwidth]{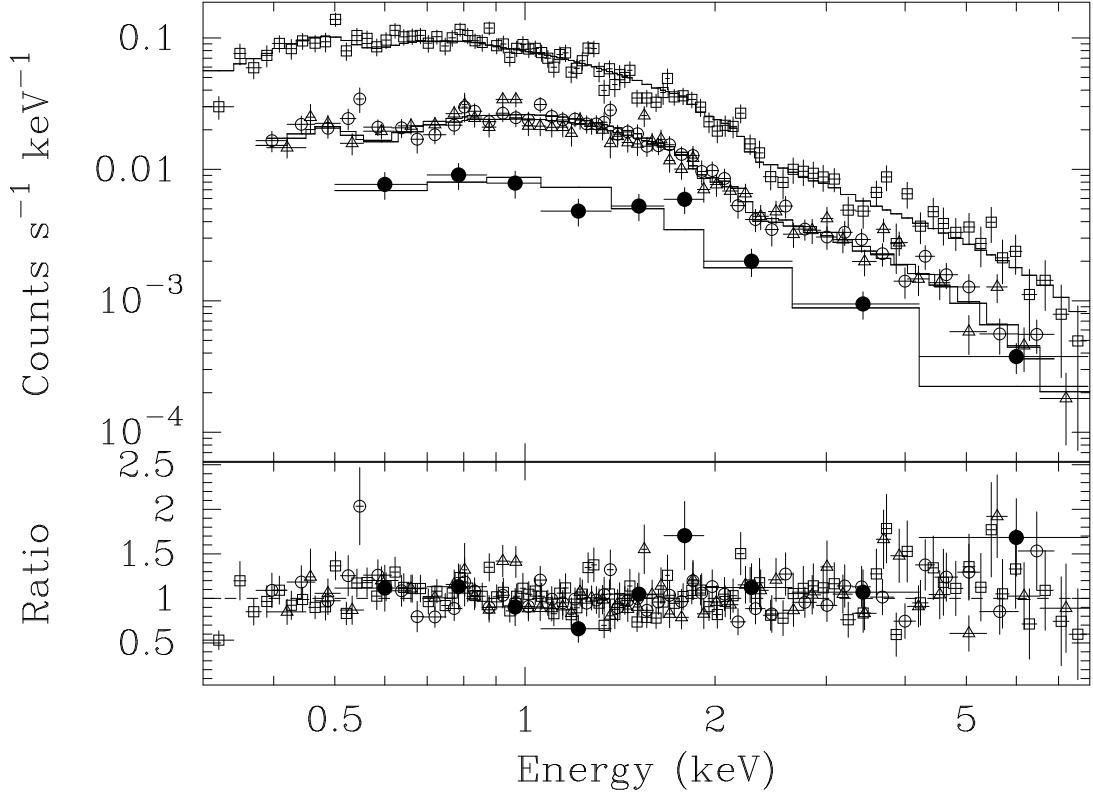} 

\caption{The spectra of the GRB\,050326 afterglow as observed by \XMM{} (PN:
squares; MOS1: open circles; MOS2: triangles) and by \Swift{}-XRT (filled
circles). The solid lines represent the best-fit absorbed power-law model
convolved with the instrumental responses (see Table~\ref{tab:XMM} for the
best-fit parameters). XRT data were selected from the time interval covered by
the \XMM{} observation.\label{fg:intercal}}
\end{figure*}

To extract the spectrum of the source in WT mode, we used the same extraction
regions, the same background regions, and the same screening as for the
temporal analysis. For PC mode, however, we further selected only grade 0--4
events in order to improve the spectral resolution. The spectrum was binned in
order to ensure a minimum of 20 counts per energy bin, ignoring channels below
0.3~keV. The spectral analysis was performed using \texttt{XSPEC} (v11.3). We
first considered WT and PC data separately, using the former for the first 2
orbits (3 to 15~ks after the burst) and the latter for the rest of the
observation. In both cases, the spectrum was fitted with an absorbed power-law
model, yielding good $\chi^2$ values (Table~\ref{tab:XRT}). The best-fit values
for the Hydrogen column density $N_{\rm H}$ and for the photon index $\Gamma$
did not show significant variations between the first (WT data) and second part
(PC data) of the observation. In fact, combining the data from the two segments
together, we obtained an excellent $\chi^2$ value, indicating that the spectral
properties of the afterglow did not change during the observation.
Fig.~\ref{fg:intercal} shows the XRT spectrum (filled circles), together with
the best-fit absorbed power-law model (precisely, only data simultaneous to the
\XMM{} observation were used for the plot; see Sect.~\ref{sec:XRT_XMM}).

To look for the presence of absorbing material in the proximity of the
afterglow, we tried to estimate the Galactic Hydrogen column density $N_{\rm
H,MW}$ towards the GRB direction. We found three different measurements:
\citet{DickeyLockman90} give $N_{\rm H,MW} = 4.6 \times 10^{20}$~cm$^{-2}$; the
Leiden/Argentine/Bonn Galactic H\,I Survey \citep{Kalberla05} provides a lower
value, $N_{\rm H,MW} = 3.8 \times 10^{20}$~cm$^{-2}$; last, the dust maps by
\citep{Schlegel98} give $A_V = 0.12$~mag, which corresponds to $N_{\rm H,MW} =
2.2 \times 10^{20}$~cm$^{-2}$ after assuming the prescription given by
\citet{PredehlSchmitt95}. While the average of these three independent
measurements is $(3.5 \pm 1.7) \times 10^{20}$~cm$^{-2}$, we conservatively
adopted the largest of the above values ($4.6 \times 10^{20}$~cm$^{-2}$).
\citet{Stratta04} estimated that the typical error affecting the maps by
\citet{DickeyLockman90} is 30\%, which is not far from the scatter  among the
three measurements. The best-fit Hydrogen column density derived by the XRT
afterglow spectrum
is marginally unconsistent with the Galactic value. Fixing $N_{\rm H} = N_{\rm
H,MW}$ provided a poor fit ($\chi^2_\nu = 1.20$ for 39 d.o.f.). The probability
of such a worsening in the fit is $< 7.5$\%, as estimated by an F-test.
Therefore, XRT data marginally suggest the presence of additional absorbing
material, likely located in the GRB rest frame. In the next section, we will
present further evidence for the presence of excess absorption, based on \XMM{}
data with better S/N ratio.

\begin{table}
\caption{Best-fit spectral parameters for the two segments of the XRT
observation (WT and PC data), fitted both separately and together. We report
the results either leaving the Hydrogen column density as a free parameter, or
freezing it to the Galactic value.\label{tab:XRT}}
\centering\begin{tabular}{lccc}\hline
            &$N_{\rm H}$ ($10^{21}$ cm$^{-2}$) &$\Gamma$               &$\chi^2_\nu$ ($\chi^2/$d.o.f.) \\ \hline
XRT (WT)    &$1.3^{+0.7}_{-0.6}$               &$1.92^{+0.27}_{-0.24}$ &1.10 $(29.7/27)$               \\
XRT (PC)    &$1.4^{+0.8}_{-0.7}$               &$2.00^{+0.38}_{-0.31}$ &1.05 $(12.6/12)$	       \\
XRT (WT+PC) &$1.4^{+0.6}_{-0.5}$               &$1.95^{+0.21}_{-0.21}$ &1.10 $(42.0/38)$	       \\ \hline
XRT (WT)    &0.46 (frozen)                     &$1.58\pm0.12$          &1.22 $(34.1/28)$	       \\
XRT (PC)    &0.46 (frozen)                     &$1.58\pm0.14$          &1.41 $(19.1/13)$	       \\
XRT (WT+PC) &0.46 (frozen)                     &$1.61\pm0.10$          &1.20 $(47.0/39)$	       \\ \hline
\end{tabular}
\end{table}

\section{\XMM{} data analysis and results}\label{sec:XMM}

The afterglow of GRB\,050326 was also observed by \XMM{} as a target of
opportunity, starting on 2005 Mar 26 at 18:25 UT (8.5~hr after the burst). The
observation lasted for 45.8~ks. Data were collected with the European Photon
Imaging Camera (EPIC), which consists of the PN
\citep{Struder01} and of the two MOS detectors \citep{Turner01}.
All the cameras were operated in full-frame mode 
with a thin and medium optical filter on PN and MOS, respectively. A
preliminary analysis of these data was presented by \citet{DeLuca_GCN}.

\subsection{Data and temporal analysis}\label{sec:XMM_temporal}

The appropriate observation data files were retrieved from the XMM Science
Archive. The data reduction was performed using the  most recent release of the
XMM Science Analysis Software (SAS v6.1.0), with the standard pipeline tasks
(\texttt{epproc} and \texttt{emproc} for PN and MOS, respectively). The
observation was badly affected by high particle background (soft proton
flares), with almost no nominal (quiescent) background  time intervals. The
back-illuminated PN CCD is particularly sensitive to this background; indeed,
more than 25\% of the PN observing time was lost due to the detector switching
to its counting mode%
\footnote{The counting mode is activated when the count rate { in a
quadrant} exceeds the telemetry limit ($\sim 400$~count~s$^{-1}$ for the PN).
In this mode, the informations for individual events of that quadrant are
not transmitted to ground.}.
The afterglow of GRB\,050326 \citep[source XMMU\,J002748.8-712217;][]{EhlePM05}
was anyway clearly detected in all cameras. The astrometry of the EPIC images
was improved by cross-correlating serendipitous X-ray sources in the field with
objects in the USNO-B1 catalog.  This yielded the following refined coordinates
for the afterglow: $\alpha_{\rm J2000}= 00^{\rm h}27^{\rm m}49\farcs1$,
$\delta_{\rm J2000} = -71\degr22\arcmin16\farcs3$, with an 1-$\sigma$
uncertainty of 1\farcs5. The EPIC and XRT positions differ by 1\farcs7,
and are therefore fully consistent within the uncertainties.

In order to retain a S/N ratio large enough to perform the temporal and
spectral analysis, a standard time-filtering approach to screen soft proton
flares could not be applied (a high particle flux was present during the whole
observation). Thus, source events were extracted with a particularly stringent
spatial selection, considering only the innermost portion of the point spread
function. We used a circle of 15\arcsec{} radius (containing $\approx 65$\% of
the EEF). The PSF correction was applied to the flux and spectral measurements
by computing the ad-hoc effective area using the SAS task \texttt{arfgen}. The
error in this procedure is estimated to be { at most at the 5\%}
level%
\footnote{{ See page~9 of\\
\texttt{http://xmm.vilspa.esa.es/docs/documents/CAL-TN-0018-2-4.pdf}}}
and it was properly taken into account in the light curve error budget.

Background events were selected from source-free regions within the same CCD
chip where the source was imaged. In particular, for the PN data we used 2
boxes of $45\arcsec \times 25\arcsec$ located at the same distance from the
readout node as the target; for the MOS we used an annulus centered at the
target position with inner and outer radii of 90\arcsec{} and 180\arcsec,
respectively. With such a choice, the background amounted to $\sim 13$\% and
$\sim 9$\% of the counts in the source extraction region for PN and MOS data,
respectively, in the 0.3--8 keV range. { The overall (background-subtracted)
number of source events was 3\,990, 1\,850 and 1\,760 in the PN, MOS1 and MOS2
detectors, respectively.

The background-subtracted count rate clearly showed a declining trend with
time. We again fitted the light curve assuming a power law decay. The value of
the decay slope $\alpha$ was evaluated independently using PN, MOS1 and MOS2
data, yielding fully consistent results. We therefore repeated the fit using
the combined dataset, finding $\alpha = 1.72 \pm 0.09$ in the 0.3--8~keV energy
range ($\chi^2_\nu = 1.30$ for 40 d.o.f.). The background-subtracted light
curve is shown in { Fig.~\ref{fig:compa}}, together with the XRT light
curve, after converting the count rates to fluxes using the best-fit absorbed
power-law models described in Sect.~\ref{sec:XRT_spectral} and
\ref{sec:XMM_spectral}).

\subsection{Spectral analysis}\label{sec:XMM_spectral}

Spectra for the source and the background were extracted from the same regions
used for the temporal analysis, as described above. Source spectra were
rebinned in order to have at least 30 counts per energy bin and to oversample
the instrumental energy resolution by a factor 3. Ad-hoc response matrices and
effective area files were created with the SAS tasks \texttt{rmfgen} and
\texttt{arfgen}, respectively. The spectral analysis was performed using
\texttt{XSPEC} (v11.3). The spectra were fitted simultaneously in the
0.3--8~keV band. Since the MOS observation started $\approx 1$~hr earlier than
the PN, a PN/MOS normalization factor was introduced in the fit as a further,
free parameter. Due to the fading of the source, this also implies that the
observed time-averaged flux is expected to be higher in the MOS than in the PN.

An absorbed power-law model reproduced the spectrum quite well ($\chi^2_\nu =
1.17$ for 190 d.o.f.). The best-fit parameter values are reported in
Table~\ref{tab:XMM} (first row); the MOS-PN normalization factor was $1.08 \pm
0.04$. Fig.~\ref{fg:intercal} shows the spectra collected by the EPIC cameras
together with the best-fit model. Both the photon index and the Hydrogen column
density are in good agreement with those found by XRT, but are much better
constrained. In particular, the value of $N_{\rm H}$ inferred from the fit was
significantly larger than the Galactic one $N_{\rm H,MW}$
(Sect.~\ref{sec:XRT_spectral}). Moreover fixing $N_{\rm H} = N_{\rm H,MW}$
resulted in a much poorer fit ($\chi^2_\nu = 1.85$ for 191 d.o.f.), and even
increasing $N_{\rm H,MW}$ by 30\% \citep[see Sect.~\ref{sec:XRT_spectral}
and][]{Stratta04} the fit was still unacceptable ($\chi^2_\nu = 1.65$ for 191
d.o.f.). A significant improvement was achieved by fixing $N_{\rm H} = N_{\rm
H,MW}$ and adding to the spectral model an extra neutral absorber at redshift
$z$ with column density $N_{{\rm H},z}$. This yielded $\chi^2_\nu = 1.03$ (189
d.o.f.); the chance  probability of such improvement, as estimated by an
F-test, was $< 1 \times 10^{-6}$ with respect to the model containing only one
absorbing component (with free $N_{\rm H}$). After adding the extra
absorption component, the best-fit power law photon index was $\Gamma =
2.03\pm0.05$, while the intrinsic gas column density and the redshift were
$N_{{\rm H},z} \sim 6 \times 10^{22}$~cm$^{-2}$ and $z \sim 6$, respectively.
However, the latter two values are not well constrained, owing to their
strong correlation (Fig.~\ref{fg:NH}). In any case, the spectral fit allowed us
to constrain $N_{{\rm H},z} > 4 \times 10^{21}$~cm$^{-2}$ and $z > 1.5$ (at
90\% confidence level for 2 parameters of interest; see inset in
Fig.~\ref{fg:NH}). We investigated the dependence of these confidence
contours on the assumed value of the Galactic column density. By varying
$N_{\rm H,MW}$ by 50\% (within the range discussed above), we found that the
90\% confidence interval on $N_{{\rm H},z}$ varies by 20\%, while that on $z$
varies by less than 10\%.

The observed (time-averaged) fluxes in the 0.2--10 keV band were $\sim 5.1
\times 10^{-13}$ and $\sim 5.5 \times 10^{-13}$~erg~cm$^{-2}$~s$^{-1}$ in the
PN and MOS, respectively. The corresponding unabsorbed fluxes were  $\sim 7.4
\times 10^{-13}$ and $\sim 7.9 \times 10^{-13}$~erg~cm$^{-2}$~s$^{-1}$.

\begin{figure}  
\includegraphics[angle=270,width=\columnwidth]{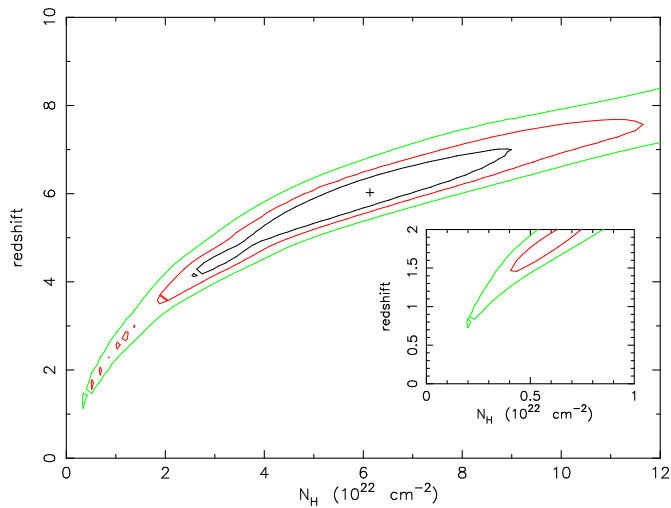} 

\caption{Confidence contours (68\%, 90\% and 99\% levels for 2 parameters of
interest) for the gas column density $N_{{\rm H},z}$ and the redshift $z$ of
the intrinsic absorber, as computed from the fit to the EPIC spectra. The
Galactic column density was assumed to be $N_{\rm H,MW} = 4.6 \times
10^{20}$~cm$^{-2}$ \citep{DickeyLockman90}. The inset shows a zoom-in of the
low-redshift region.\label{fg:NH}}
\end{figure}

In order to search for possible line features in the spectrum (both in emission
and in absorption) we divided the 0.4--5~keV energy range into 0.2~keV
intervals. For the continuum, we assumed the best-fit absorbed power-law model
(including also the rest-frame column density). For each of the intervals, we
added a Gaussian line with fixed width (smaller than the instrumental
resolution) and central energy free to vary within the selected interval; the
normalization could be either positive or negative. We then repeated the
exercise with different choices of the energy intervals. We found no
significant lines in the 0.4--5~keV range in the combined MOS/PN dataset. The
upper limit (3-$\sigma$) on the equivalent width of any line is $\sim 50$ and
$\sim 250$~eV in the 0.4--2 and 2--5~keV energy ranges, respectively.

We also tried to fit the spectra with thermal models
\citep[e.g.][]{Reeves02,Lazzati03}. A redshifted, optically thin plasma
emission  model (MEKAL in \texttt{XSPEC}) was used, with the redshift linked to
that of the intrinsic absorber. The fit worsened ($\chi^2_\nu \approx 1.4$)
with respect to the simple power law model, either fixing the metal abundances
to Solar values, or leaving them as free parameters.

Finally, we looked for possible spectral evolution with time,  using the power
law plus redshifted absorber model described above. For this study we divided
the data into two subsets with exposure times of $\sim 15.8$ and $\sim
27.6$~ks, each subset containing approximately half of the afterglow counts. We
then extracted the corresponding spectra for the source and the background. No
significant ($>3\mbox{-}\sigma$) variations in the spectral parameters were
found (except, of course, for the flux normalization).

\begin{table}

\caption{The spectral parameters as measured separately by \XMM{} and XRT,
fitting the data both separately and together. For this fit, XRT data were
selected from the same time interval covered by the \XMM{} observation. Due to
the limited statistics, $N_{\rm H}$ was frozen to the value derived by \XMM{}
when fitting XRT data. The errors are at 90\% confidence level for a single
parameter of interest.\label{tab:XMM}}

\centering\begin{tabular}{lccc}\hline
Instrument  &$N_{\rm H}$ ($10^{21}$ cm$^{-2}$) &$\Gamma$                &$\chi^2_\nu$ ($\chi^2/$d.o.f.) \\ \hline
EPIC        &$1.3^{+0.1}_{-0.1}$               &$2.13^{+0.06}_{-0.06}$  &1.17 (222.5/190)               \\
XRT         &1.3 (frozen)                      &$1.96^{+0.29}_{-0.27}$  &1.43 (11.4/8)                  \\
EPIC + XRT  &$1.3^{+0.1}_{-0.1}$               &$2.09^{+0.05}_{-0.08}$  &1.18 (219.7/197)               \\ \hline
\end{tabular}

\end{table}

\section{\XMM{}/\Swift-XRT comparison}\label{sec:XRT_XMM}
As noted above, the spectral parameters as derived using XRT and \XMM{} data
nicely agree within the errors. To better check the consistency between the
afterglow temporal and spectral properties as derived by the two satellites, we
performed a more accurate operation. 
We selected XRT data from the MOS observing time interval (35--76~ks after the
burst). During this period, the \XMM{} observation was continuous, whereas
\Swift{} completed eight orbits, providing 13~ks of effective exposure time for
XRT. Therefore, in order to compare the two datasets, we had to assume that
both the spectrum and the light curve behaved in a regular fashion. For
example, a flare during a \Swift{} occultation would distort the results of the
comparison. This assumption seems anyway fully justified by the \XMM{} data,
which show a regular behaviour of the light curve without significant
irregularities.

We fitted XRT data using the absorbed power law model. For the sake of
comparison between the two instruments, we did not add any rest-frame
absorption component. Due to the limited statistics, moreover, we froze the
column density to the value found by \XMM{} (which is fully consistent with
that found by XRT for the full observation). The results of the fit are
presented in Table~\ref{tab:XMM}. Within the errors, the photon index is
consistent with that found by \XMM{}. Last, we fitted together the XRT and
\XMM{} data from the three EPIC instruments (leaving $N_{\rm H}$ as a free
parameter), obtaining our final best fit (Table~\ref{tab:XMM}).

Our second step was to compare the flux normalization factors for the two
instruments. To this extent, we selected the data  only from exactly
overlapping intervals (8.5~ks effective observation time).
We froze $N_{\rm H}$ and $\Gamma$ to the values found previously (which rest on
a better statistics), computing only the normalization factors (the ratio of
the fluxes observed by \Swift-XRT and the EPIC detectors). The fit provided
$0.90 \pm 0.14$ for MOS1, $0.95 \pm 0.16$ for MOS2, and $0.89 \pm 0.15$ for PN.
This result indicate that, within the errors, the XRT current absolute flux calibration
is good, providing perhaps slightly underestimated values (at the $\sim 5$\%
level).

As shown in the previous sections, the light curves from XRT and \XMM{} were
well described by power laws with slopes $1.64\pm 0.07$ and $1.72 \pm 0.09$,
respectively. These values are consistent within their errors. Moreover,
selecting XRT data from the 8.5~ks time interval with simultaneous \XMM{}
observations, we found a slope $\alpha = 1.73 \pm 0.12$, in perfect agreement
with the \XMM{} result (and consistent with that measured by XRT for the whole
observation).

In order to directly compare the fluxes measured from the two satellites, we
converted the 0.3--8 keV count rates to unabsorbed fluxes in the same band
(Fig.~\ref{fig:compa}). To compute the conversion factor for the \XMM{}
spectrum, we used the absorbed (one component) power law model with the
parameters reported in Table~\ref{tab:XMM} (first row). For XRT, we calculated
a conversion factor for each of the two operational modes, using in both cases
the absorbed power law best-fit model (leaving $N_{\rm H}$ as a free
parameter), as reported in Table~\ref{tab:XRT} (first and second rows). The
combined fit of the joint XRT and XMM data provided $\alpha = 1.70 \pm 0.05$
(Fig.~\ref{fig:compa}). In the following, we will adopt this value as the best
determination of the temporal decay slope.

\section{Optical and ultraviolet observations}\label{sec:UVOT}

\begin{table*}

\caption{Summary of optical and ultraviolet observations. All data but the
point in the $R$ band were measured by \Swift-UVOT. The { $R$-band}
observations is by \citet{Tristram05}. All measurements were corrected for the
Galactic extinction \citep[$A_V = 0.123$~mag;][]{Schlegel98}, assuming the
Milky Way extinction curve by \citet{Pei92}. The optical-to-X-ray spectral
index is $\beta_{\rm OX} = \log(F_\nu^{\rm opt}/F_\nu^{\rm X}) / \log(\nu_{\rm
X}/\nu_{\rm opt})$. The X-ray frequency was set to $\nu_{\rm X} = 1.5$~keV, the
logarithmic mean of the XRT observing range. The X-ray flux was computed by
interpolating the XRT light curve { to} the time of the optical limit. The
last column reports the ratio of the extrapolated X-ray flux to the optical one
(assuming no spectral breaks between the two bands; see
Fig.~\ref{fg:SED}).\label{tab:optical}}

\centering\begin{tabular}{lllllllll} \hline
\bf Time since burst &\bf Filter &\bf Wavelength  &\bf Exposure time &\bf Extinction &\bf Magnitude &\bf Flux density &$\beta_\mathbf{OX}$ & \bf X-ray/optical ratio\\
(s)                  &           &(\AA) 	  &(s)  	     & (mag)         &              &($\mu$Jy)       &		&                    \\ \hline
3300                 &UVW2       &1930  	  &10		     &0.29           &$> 17.41$     &$< 99.11$	     &$< 0.52$  &$> 23^{+8.9}_{-6.4}$\\
3306                 &$V$        &5460  	  &100  	     &0.12           &$> 18.66$     &$< 107.5$	     &$< 0.45$  &$> 65 ^{+31}_{-21}$ \\
3350                 &UVM2       &2220  	  &100  	     &0.36           &$> 18.54$     &$< 34.17$	     &$< 0.32$  &$> 75 ^{+30}_{-21}$ \\
3470                 &UVW1       &2600  	  &100  	     &0.27           &$> 18.75$     &$< 27.80$	     &$< 0.29$  &$> 103^{+43}_{-30}$ \\
5219                 &$U$        &3450  	  &750  	     &0.20           &$> 19.65$     &$< 11.00$	     &$< 0.24$  &$> 176^{+77}_{-54}$ \\
9390                 &$B$        &4350  	  &714  	     &0.16           &$> 20.87$     &$< 17.47$	     &$< 0.46$  &$> 52 ^{+24}_{-16}$ \\ \hline
24840             &$R_{\rm MOA}$ &6399  	  &600  	     &0.10           &$> 20.20$     &$< 25.70$	     &$< 0.74$  &$> 10^{+5.0}_{-3.4}$\\ \hline
\end{tabular}
\end{table*}

The UVOT instrument onboard \Swift{} observed the field of GRB\,050326 together
with XRT, starting 54~min after the trigger. In the subsequent orbits, it
collected a series of images in its 6 broad-band filters ($V$, $B$, $U$, UVW1,
UVM2, and UVW2; Table~\ref{tab:optical}). The afterglow was not detected in any
of the single or coadded exposures. Summing the images in each of the six
filters, we estimated the 3-$\sigma$ upper limits using the UVOT dedicated
software (task \texttt{uvotsource}). The counts were extracted from a
6\arcsec{} and 12\arcsec{} radius aperture for the optical and ultraviolet
filters, respectively, after subtracting the background.
We then corrected the upper limits for Galactic absorption \citep{Schlegel98},
assuming the extinction curve of \citet{Pei92}. Our final limits are summarized
in Table~\ref{tab:optical}. With respect to the original values reported by
\citet{Holland05}, our measurements were obtained adopting the most recent
on-flight calibration.

The only reported ground-based optical observation for this burst was an
$R$-band upper limit provided by the 0.6m telescope at the Mt. John Observatory
\citep{Tristram05}. This measurement is also listed in Table~\ref{tab:optical}.

\section{Discussion}\label{sec:discussion}

\subsection{The X-ray light curve}\label{sec:lightcurve}

To date, \Swift{} has observed X-ray emission from dozens of GRB afterglows. A
systematic analysis of their light curves has evidenced several common features
\citep{Nousek05,Chinca05}. During the first few hundred seconds, a steep decay
is often observed \citep[$\alpha \approx 3\mbox{--}5$;][]{Taglia05}, usually
interpreted as the tail emission from the prompt GRB
\citep[e.g.][]{Cusu_050319}. This phase is followed by a much flatter decline
\citep[$\alpha\approx0$--0.7; e.g.][]{Campana05,Nousek05}, lasting up to
$10^3\mbox{--}10^5$~s (and in some cases even longer). Then, the light curve
steepens again, leading to $\alpha \approx 1$--$1.5$; this phase was the one
seen by \textit{Beppo}\/SAX, { \XMM{} and \textit{Chandra}}.
At late times, a further steepening is sometimes
observed \citep[e.g.][]{Vaughan05}, likely the signature of a jetted outflow
\citep{Rhoads99}. In some cases, bumps and flares appear superimposed to the
power-law decay, up to several tens of ks \citep[e.g.][]{Burrows05}.

The light curve of GRB\,050326 exhibited a different behaviour with respect to
that outlined above. 
Its light curve showed a single, unbroken decay from $\approx 55$~min to
$\approx 4.2$~d. However, our coverage began relatively late, so that we may
have missed early deviations from the power law behaviour.

In order to investigate the afterglow early stages and to analyze the
connection between the prompt and afterglow emission, we extrapolated the
afterglow flux to the time of the prompt emission. We then compared the
obtained value with that expected from the prompt emission in the XRT band
(0.3--8~keV), computed adopting the Band best-fit model. The result is shown in
Fig.~\ref{fig:compa}, where the light circles indicate the prompt emission
fluxes. Since the GRB spectrum is known in good detail (particularly, since no
breaks are expected between the BAT and XRT ranges), the extrapolation process
should be quite reliable. As it can be seen, 
if no temporal breaks were present in the X-ray light curve, the afterglow
flux in the X-ray range exceeded the prompt one by a factor of $\sim 100$ 
(with a small uncertainty, due the tiny error in the decay index).
We cannot exclude that such emission was present (since we have no prompt
observations in the X-ray band), but, if present, the present component would
appear as a very bright, soft excess. Such feature would not be unprecedented
\citep{Vanderspek04,Vetere05}, but in this case it would likely contaminate the
low-frequency end of the BAT spectrum. Moreover, the soft excesses always
contained less energy that the GRB proper. Thus, the most conservative
hypothesis is to assume that a break was present in the early light curve, or
that the afterglow onset was delayed. Indeed, as mentioned above, most of
\Swift{} afterglows show a shallow decline phase during the first thousands
seconds after the GRB. Independently of any extrapolation, we note that
GRB\,050326 was distinctly different from most bursts observed by
\textit{Beppo}SAX, for which the backward extrapolation of the late-time X-ray
afterglow roughly matched the prompt emission level in the X-ray range, as
measured by the Wide Field Cameras \citep{Frontera00}.

We also performed a different operation. Using the best-fit X-ray spectrum, we
extrapolated the XRT flux to the BAT energy range (20--150~keV), and reported
it at the time of the burst using the afterglow decay law. Also in this case,
the expected value exceeded the observed prompt emission, but by a smaller
factor. This again suggests that a break in the light curve was present before
the beginning of the XRT observation, but the evidence is less compelling. For
example, we cannot even exclude that the afterglow spectrum had a break between
the XRT and BAT ranges, so that the extrapolation actually overestimated its
flux.

\subsection{Constraints on the afterglow parameters}\label{sec:models}

The properties of the explosion can be inferred in the context of the standard
afterglow model \citep[e.g.][]{MeszRees97,Sari98}. In this context, the
observed emission is due to synchrotron radiation from a decelerating
relativistic shock, which produces a decaying flux with a power-law spectrum.
Depending on the model parameters, definite relations between the spectral and
temporal indices $\alpha$ and $\beta$ are predicted. The combined \XMM{} and
\Swift-XRT data provide $\alpha = 1.70 \pm 0.05$ and $\beta = 1.09 \pm 0.08$.
Both values are not unusual among GRB afterglows at comparable epochs
\cite[e.g.][]{DePasq05,Chinca05,Nousek05}. These numbers are consistent with a
spherical outflow expanding inside a homogeneous medium, if the XRT range was
between the injection and cooling frequencies ($\nu_{\rm i}$ and $\nu_{\rm c}$,
respectively). In this case, the model prediction is $\alpha = 3\beta/2 = 1.63
\pm 0.12$, in excellent agreement with the measured value $\alpha = 1.70 \pm
0.05$. All other possibilities (a wind-stratified medium, or a different
location of the break frequencies) are excluded at $> 3.5$-$\sigma$ level. The
power-law index of the electron energy distribution is $p = 1 + 4\alpha/3 = 1 +
2\beta$, so that $p = 3.25 \pm 0.06$. Such value is rather high, but not
unprecedented.

No break was observed in the X-ray light curve of GRB\,050326 between 55~min
and $\sim 4.2$~d after the burst. The condition $\nu_{\rm i} < \nu < \nu_{\rm
c}$ thus held during this time range. While $\nu_{\rm i}$ typically lies below
$\sim 10^{15}$~Hz for $t > 1$~hr \citep[e.g.][]{Sari98,PanKum00}, keeping $\nu
< \nu_{\rm c}$ up to $t > 4.2$~d { requires $n_0 \varepsilon_{\rm
B,-2}^{3/2} < 3 \times 10^{-5} E_{\rm iso,54}^{-1/2}$}, where $n = n_0 \times
1$~cm$^{-3}$ is the ambient particle density, $\varepsilon_{\rm B} = 10^{-2}
\varepsilon_{\rm B,-2}$ is the magnetic field energy fraction, and $E_{\rm iso}
= E_{\rm iso,54} \times 10^{54}$~erg is the (isotropic-equivalent) fireball
energy. This condition is difficult to satisfy
\citep[e.g.][]{PanKum01,Yost03}, so it may be regarded as a problem for the
model. We note, however, that for this burst both $\alpha$ and $\beta$ could be
measured with good accuracy, so the consistency between the predicted and
observed value of the decay index is remarkable.

\begin{figure*}
\centering\includegraphics[width=0.7\textwidth]{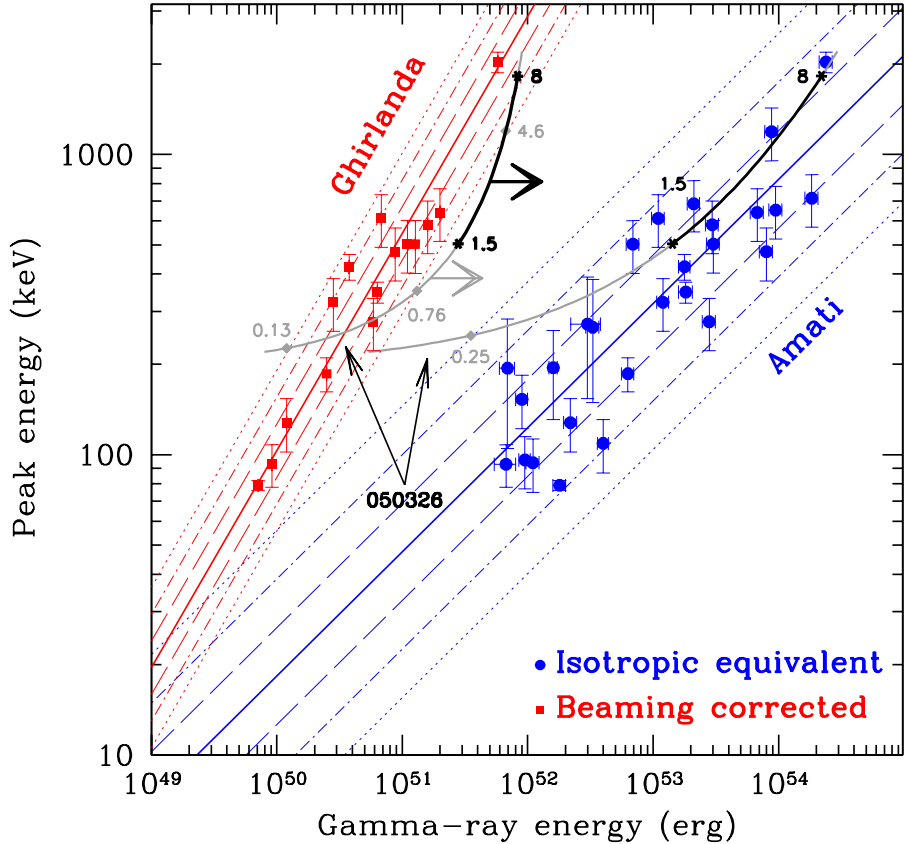}

\caption{Comparison of GRB\,050326 with the Amati (right) and Ghirlanda (left)
relations \citep{Amati02,Ghirla04}. The thick solid curves (black and grey)
show the position of GRB\,050326 as its redshift varies in the interval $0.1 <
z < 10$. The Ghirlanda track is actually a boundary (as the horizontal arrows
indicate), since we can infer only a lower limit to the beaming-corrected
energy at each redshift. Filled circles and squares indicate the GRBs which
define the above two relations, plotted as straight solid lines (together with
their 1-, 2- and 3-$\sigma$ contours: long-dashed, dot-dashed and dotted lines,
respectively). Data were taken from \citet{Ghirla04,Ghirla05}. Grey diamonds
indicate the intersection of the GRB\,050326 tracks with the 3-$\sigma$
contours of the Amati and Ghirlanda relations. These points thus define the
3-$\sigma$ redshift ranges for which GRB\,050326 was consistent with the two
relations. In the two GRB\,050326 tracks, the region $1.5 < z < 8$ (indicated
by the X-ray data) is shown in black, bound by asterisks.\label{fg:Ghirlanda}}

\end{figure*}

The absence of any break also poses some constraints on the geometry of the
emission. GRB afterglow light curves often show a late-time steepening,
commonly interpreted as the result of a jetted geometry. For GRB\,050326, this
break likely occurred \textit{after} the end of the \Swift ~observations. In
fact, breaks earlier than $\sim 1$~hr are usually due to different reasons
\citep[such as the end of the refreshed shock episode; e.g.][]{Zhang05}. If
interpreted as a jet break, a very narrow jet would be implied. Moreover, the
decay slope in the monitored time range is quite flat compared with that
expected (and usually observed) after jet breaks
\citep[e.g.][]{Israel99,Harrison01,Klose04}. The measured decay would imply a
hard electron distribution ($p < 2$). Using the relations provided by
\citet{DaiCheng01}, an unreasonably low $p = 4\alpha - 6 = 0.8 \pm 0.2$ would
result (with the X-ray band being above $\nu_{\rm c}$), which would give rise
to a spectrum completely inconsistent with the observed one. Assuming $\nu_{\rm
c}$ above the observed range would only worsen the situation.

Assuming that the jet break occurred at $t_{\rm b} \ga 4$~d allows us to put
some constraints on the jet opening angle and on the beaming-corrected
gamma-ray energy. To this extent, the main obstacle is the lack of a well
constrained redshift. Our spectral analysis of the \XMM{} data allowed only to
set a broad range $1.5 < z < 8$ (90\% confidence level). Using the bolometric
fluence $\mathcal{F}$ of the prompt emission, the lower limit on $z$ provides a
constraint on the GRB radiated energy:
\begin{equation}
  E_{\gamma,\rm iso} = 4\pi \frac{D_{\rm L}^2(z)}{1+z} \mathcal{F} >
  1.4 \times 10^{53}~\mathrm{erg},
\end{equation}
where $D_{\rm L} > 3.40 \times 10^{28}$~cm is the luminosity distance.
GRB\,050326 was therefore quite likely bright in gamma rays compared to other
GRBs detected by \Swift{} \citep[e.g.][]{Chinca05}.

Following \citet{SariPiranHalpern}, the jet half-opening angle can be
constrained as follows:
\begin{equation}
  \vartheta_{\rm j} > 6.7\degr \left( \frac{1+z}{2.5} \right)^{-3/8}
    \left( \frac{E_{\gamma\rm,iso}}{10^{53}~\mathrm{erg}} \right)^{-1/8}
    \left( \frac{\eta}{0.2} \frac{n}{1~\mathrm{cm}^{-3}}\right)^{1/8},
\end{equation}
where $\eta$ is the prompt radiative efficiency and we have assumed $t_{\rm b}
> 4$~d. Note that the dependance from the gamma-ray energy and on the other
parameters is rather mild. The corresponding beaming-corrected gamma-ray energy
is then
\begin{equation}
  E_{\gamma,\rm j} = \frac{\pi}{2} \vartheta_{\rm j}^2 E_{\gamma,\rm iso}
    \propto \left( \frac{E_{\gamma,\rm iso}}{1+z} \right) ^{3/4}
    \propto \mathcal{F}^{3/4} \left( \frac{D_{\rm L}}{1+z}\right)^{3/2}.
\end{equation}
The quantity $D_{\rm L}/(1+z)$ has only a mild dependence upon $z$. For the
redshift range allowed by our X-ray measurements ($1.5 < z < 8$), we have $1.4
\times 10^{28}~\mathrm{cm} < D_{\rm L}/(1+z) < 2.8 \times 10^{28}~\mathrm{cm}$,
so that $E_{\gamma,\rm j} > (3\mbox{--}8) \times 10^{51}$~erg. This is at the
high end of the beaming-corrected gamma-ray energy distribution
\citep{Ghirla04}.

Fig.~\ref{fg:Ghirlanda} shows the position of GRB\,050326 in the plane
$E_\gamma$ vs $E_{\rm p}$, to check how it compares with the Amati and
Ghirlanda relations \citep{Amati02,Ghirla04}. As the redshift varies, the burst
position follows the tracks indicated by the thick solid curves (see the figure
caption for the details). The Ghirlanda track is actually a boundary, since the
lower limit on $t_{\rm b}$ translates into a lower limit for $E_{\gamma,\rm j}$
for any given redshift. We can ask for which redshifts GRB\,050326 was
consistent with the two above relations. Only loose limits are provided by the
Amati relation (which has a large dispersion): at the 3-$\sigma$ level, $z >
0.25$ is implied. The comparison with the Ghirlanda relation is less solid,
since further assumptions are needed (such as the ambient particle density and
the break time). However, for our fiducial values, we have two allowed ranges:
a low-redshift region ($0.1 \la z \la 0.8$), plus a high-redshift solution ($z
\ga 4.5$). A lower particle density would move the track towards the left,
while a larger jet break time would shift it rightwards.

We note that the only way to make GRB\,050326 in agreement with the Ghirlanda
relation and simultaneously satisfy our constraints on the X-ray absorption ($z
\ga 1.5$) is to require a high redshift for this event, since the low-redshift
region is excluded by the fit to the X-ray column density at $> 99$\%
confidence level. Therefore, although our arguments are rather speculative, and
would surely need more conclusive data, we regard GRB\,050326 as a
moderate/high redshift candidate.

Recently, \citet{LiangZhang05} have presented a model-independent
multidimensional correlation between the observed isotropic energy
$E_{\gamma,\rm iso}$, the rest-frame spectral peak energy $E_{\rm p}$, and the
comoving break time $t_{\rm b}/(1+z)$. This relation, in principle, allows to
compute $E_{\gamma,\rm iso}$ (and $z$) for a GRB with known $E_{\rm p}$ and
$t_{\rm b}$. However, no significant constraints could be inferred in the case
of GRB\,050326. Moreover, since this relation was derived using the break time
as measured in the optical band, its application to X-ray data may not be
valid \citep{LiangZhang05}.

\subsection{Evidence for intrinsic absorption}\label{sec:absorption}

The presence of intrinsic absorption, besides allowing us to constrain the GRB
redshift, has other important consequences. The rest-frame absorbing system has
a Hydrogen column density larger than $\sim 4 \times 10^{21}$~cm$^{-2}$. For
moderate redshifts, $N_{{\rm H},z}$ would be much larger. Several afterglow
observations, both from \Swift{} and previous missions, showed evidence for
excess X-ray absorption in addition to the Galactic value
\citep[e.g.][]{GalamaWijers01,Stratta04,DeLuca05}. Recently, \citet{Campana_NH}
showed that about half of \Swift{} afterglows have a large rest-frame column
density, typical of giant molecular clouds \citep{ReichartPrice02}. Given the
connection between GRB explosion and supernovae
\citep{Galama98,Stanek03,Hjorth03}, this fact may constitute a powerful way to
study the regions where massive star formation takes place in the high-redshift
Universe. \citet{LazzatiPerna02} showed that the prompt GRB flux is able to
ionize the surrounding medium up to radii as large as $\sim 5$~pc, therefore
leaving no absorbing material. Such process may have been observed in act for
GRB\,000528 \citep{Frontera04}. The fact that a large column density was
measured in GRB\,050326 may imply that the absorbing material was distributed
in a wide region ($R \ga 5$~pc), or that the ionizing flux was not large.

\begin{figure}  
\includegraphics[width=\columnwidth]{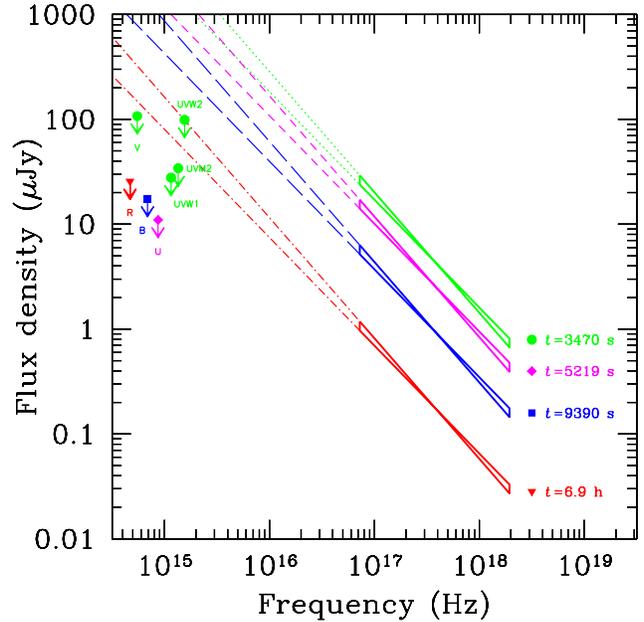}

\caption{Broad-band spectral energy distribution of the afterglow of
GRB\,050326, computed at different times (which are identified by different
symbols). The shape of the X-ray spectrum was assumed to be constant throughout
the observation, and the decay law was adopted to report the X-ray flux at the
time of the optical measurements.\label{fg:SED}}

\end{figure}

For GRB afterglows, comparable absorption in the X-ray range is usually not
accompanied by large extinction in the optical band \citep{GalamaWijers01}.
Only small amounts of dust are usually inferred from the analysis of optical
spectra, even when heavy extinction is observed in the X-ray afterglow. Several
explanations were invoked to explain this discrepance, among which the
destruction of dust from the burst and/or afterglow photons
\citep{WaxmanDraine00}, a large gas-to-dust ratio in the intervening
material \citep{Stratta04}, or an overabundance of $\alpha$ elements
\citep{Watson05}. Unluckily, several factors hamper the study of this problem,
such as the uncertaintes in the shape and normalization of the extinction
curve, the possibility that GRBs occur in special, low-metallicity environments
\citep{Fynbo03,McFadyenWoosley99}, and in several cases the lack of the
redshift determination.

For GRB\,050326, no detection in the optical/ultraviolet band could be
obtained. Table~\ref{tab:optical} reports the available upper limits to the
afterglow flux, from both \Swift-UVOT and ground-based observations. In
Fig.~\ref{fg:SED} we show the optical-to-X-ray spectral energy distribution at
different epochs. We computed the X-ray flux at the time of each available
limit, adopting the decay law measured by \Swift-XRT and \XMM. Furthermore, no
spectral evolution was assumed.

As it can be seen, the UVOT limits provide strong constraints, even if they
are not particularly deep. In fact, GRB\,050326 was bright in X rays (with a
flux of $\approx 10^{-10}$~erg~cm$^{-2}$~s$^{-1}$, 1~hr after the GRB).
Table~\ref{tab:optical} reports the optical-to-X-ray spectral indices
$\beta_{\rm OX}$. For all our measurements, $\beta_{\rm OX} < \beta_{\rm X}$.
Moreover, most of them violate the synchrotron limit $\beta_{\rm OX} > 0.5$,
which holds for a non-obscured synchrotron spectrum. GRB\,050326 can therefore
be classified as truly dark, according to the definition proposed by
\citet{Jakob04}. This limit is quite robust, since no assumptions are made
about the position of the synchrotron break frequencies. For this GRB, however,
we can go further in this reasoning. Our analysis of the temporal and spectral
properties of the afterglow, in fact, indicated that the XRT range was below
the cooling frequency $\nu_{\rm c}$. Therefore, the extrapolation of the XRT
spectrum to the optical domain seems in this case reliable, since no spectral
breaks are expected to lie between these two bands. This allows estimating the
suppression factor suffered by the optical flux, and is reported in the last
column of Table~\ref{tab:optical}. Again, large lower limits were found,
implying conspicuous rest-frame extinction (up to a factor $\sim 100$ and more,
corresponding to $> 5$~mag). The presence of the injection frequency $\nu_{\rm
i}$ close to or blueward of the optical band may partly explain the flux
dearth. However, following the formulation of \citet{PanKum00}, even choosing
rather extreme parameters ($E_{\rm iso} = 10^{54}$~erg, $z = 5$,
$\varepsilon_{\rm e} = \varepsilon_{\rm B} = 0.1$), $\nu_{\rm i}$ can at most
be comparable to the ultraviolet observed frequencies. In particular, at the
time of the $U$-filter measurement, which provides the strongest constraint,
$\nu_{\rm i}$ cannot be blueward of this band. So, even if $\nu_{\rm i}$ has
some role in this game, it cannot be responsible for the whole suppression of
the optical flux. Moreover, low values of $\varepsilon_{\rm B}$ were required
to keep the cooling frequency outside the XRT range (see
Sect.~\ref{sec:models}), so that $\nu_{\rm i}$ was likely at much lower energies
than the optical band.

The truly dark nature of this burst allows one of the following two
possibilities. The burst may have suffered dust extinction in its host galaxy.
The amount of dust is not straightforward to evaluate. The main obstacle is
again the lack of the redshift, together with the unkwown shape of the
extinction curve. However, our limit that more than 5~mag were missing in the
observed $U$ band may roughly correspond to $A_V \ga 2$~mag for $z \sim 1.5$,
 even if many other solutions are acceptable}. The second possibility is
that GRB\,050326 was at high redshift, as suggested by our analysis of the
X-ray spectrum combined with the limits provided by the Ghirlanda relation. In
this case, virtually no flux is left blueward of the redshifted Lyman dropout.
To suppress the flux in the $V$ band, $z \ga 5$ would be required. However, the
combination of a moderate redshift and mild absorption may relax this
condition.

\section{Conclusions}\label{sec:conclusion}

We have presented a detailed analysis of the GRB\,050326 prompt and afterglow
emission. The combined capabilities of \Swift{} (which sampled the light curve
for a relatively long time span) and \XMM{} (which ensured a large statistics)
allowed to obtain a thorough characterization of the afterglow properties.

The prompt emission was relatively bright (with a 20--150~keV fluence of $\sim
8 \times 10^{-6}$~erg~cm$^{-2}$). The spectrum was hard (photon index $\Gamma =
1.25 \pm 0.03$), suggesting a peak energy at the high end of the BAT energy
range or beyond. Indeed, thanks to the simultaneous detection of this burst by
the \textit{Wind}-Konus experiment \citep{Golenetskii05}, the prompt spectrum
could be fully characterized. The prompt bolometric fluence was $\mathcal{F}
\sim 2.4 \times 10^{-5}$~erg~cm$^{-2}$ (1--10\,000~keV), and the observed peak
energy was $E_{\rm p,obs} = 200 \pm 30$~keV.

Due to pointing constraints, XRT and UVOT observations could start only 54~min
after the GRB. The X-ray afteglow was quite bright, with a flux of $7 \times
10^{-11}$~erg~cm$^{-2}$~s$^{-1}$ (0.3--8~keV) 1~hr after the GRB. However, no
optical counterpart could be detected. The X-ray light curve showed a steady
decline, with no breaks or flares. The best-fit power-law decay index was
$\alpha = 1.70 \pm 0.05$. Such regular behaviour is different from that usually
observed by \Swift, but this may be the result of the limited time coverage
(observations could be carried out only between 54~min and 4.2~d after the
burst). Indeed, extrapolation of the afterglow light curve to the time of the
prompt emission overpredicts the burst flux, and may suggest a slower decay
before the beginning of the XRT observation.

The analysis of the combined XRT and \XMM{} data allowed to characterize in
detail the afterglow spectrum. A fit with an absorbed power-law model provided
a good description to the data, yielding a photon index $\Gamma = 2.09 \pm
0.08$ and a column density significantly in excess to the Galactic value. The
best-fit model was thus computed adding an extra absorption component, leaving
its redshift $z$ free to vary. Although both $N_{{\rm H},z}$ and $z$ could not
be effectively constrained, a firm lower limit $N_{{\rm H},z} > 4 \times
10^{21}$~cm$^{-2}$ could be set. Therefore, GRB\,050326 adds to the growing set
of afterglows with large rest-frame column density
\citep{GalamaWijers01,Stratta04,Campana_NH}. The limits measured in the optical
and ultraviolet region by UVOT lie well below the extrapolation of the X-ray
spectrum. In particular, they violate the synchrotron limit that the
optical-to-X-ray spectral index should be larger than 0.5. This implies a large
extinction and/or a high redshift.

The X-ray spectral analysis also allowed us to set the lower limit $z > 1.5$ to
the redshift of the absorbing component (and therefore of the GRB). The
isotropic-equivalent gamma-ray energy was then $E_{\gamma,\rm iso} > 1.4 \times
10^{53}$~erg. The temporal and spectral properties of the afterglow were nicely
consistent with a spherical fireball expanding in a uniform medium, with the
cooling frequency above the X-ray range. We could therefore set a lower limit
to the jet break time $t_{\rm b} \ga 4$~d. The jet opening angle could be
constrained to be $\vartheta_{\rm j} \ga 7\degr$, with only a weak dependence
on the (unknown) fireball energy. The beaming-corrected gamma-ray energy was
$E_{\gamma,\rm j} = (3\mbox{--}8) \times 10^{51} (t_{\rm b}/4~{\rm
d})^{3/4}$~erg, independently from the redshift. GRB\,050326, thus, released a
large amount of gamma rays \citep[only GRB\,990123 had a larger
energy in the sample of][]{Ghirla04}.

To be consistent with the Ghirlanda relation \citep{Ghirla04}, two redshift
ranges are allowed, either at low ($z \la 0.8$) or high ($z \ga 4.5$) redshift.
However, to simultaneously satisfy also the limits derived from the X-ray
spectral analysis, only the high-redshift region is left. We note that the
Ghirlanda relation is still based upon a small sample, so that any inference
cannot yet be regarded as conclusive. However, the results from the X-ray
spectra, the consistency of the GRB\,050326 properties with the Ghirlanda
relation, and the strong dearth of optical/ultraviolet afterglow flux, are
overall consistent with a moderate/high redshift ($z \ga 4$). A search for the
host galaxy through deep infrared and optical imaging may conclusively settle
this issue.

\begin{acknowledgements}

This work is supported at OAB by the ASI grant I/R/039/04, at Penn State by the
NASA contract NAS5-00136, and at the University of Leicester by the PPARC grant
PPA/Z/S/2003/00507. We gratefully acknowledge the contributions of dozens of
members of the XRT and UVOT teams at OAB, PSU, UL, GSFC, ASDC, and MSSL, and
our subcontractors, who helped make this instrument possible. DM thanks INAF
for financial support.

\end{acknowledgements}

\end{document}